\documentclass[11pt]{article}
\usepackage[utf8]{inputenc}
\usepackage[english]{babel}
\usepackage{amsmath,amsfonts,amsthm,graphicx,float}
\usepackage{natbib}

\DeclareMathAlphabet{\mathpzc}{OT1}{pzc}{m}{it}

\usepackage{algorithm}

\usepackage[noend]{algpseudocode}

\def\b1{{\mathbf 1}}

\usepackage[font=small]{caption}

\usepackage{titlesec}

\titleformat*{\section}{\normalfont\fontsize{14}{17}\bfseries}
\titleformat*{\subsection}{\normalfont\fontsize{12}{15}\bfseries}

\usepackage{bm}
\usepackage{enumerate}

\usepackage{soul}
\usepackage[usenames, dvipsnames]{color}

\date{}

\begin{document}

\title{\LARGE Nonparametric geostatistical risk mapping}\normalsize
\author{Rub\'en Fern\'andez-Casal \\
Universidade da Coru\~{n}a\thanks{%
Research group MODES, CITIC, Department of Mathematics, Faculty of Computer Science, Universidade da Coru\~na, Campus de Elvi\~na s/n, 15071,
A Coru\~na, Spain}
\and
Sergio Castillo-P\'aez \\
Universidad de las Fuerzas Armadas ESPE\thanks{Departamento de Ciencias Exactas, Universidad de las Fuerzas Armadas ESPE, Av. General Rumi\~nahui s/n, 171103, Sangolqu\'i, Ecuador}
\and %
Mario Francisco-Fern\'andez\\
Universidade da Coru\~{n}a\footnotemark[1]
}
\maketitle


\begin{abstract}
In this work, a fully nonparametric geostatistical approach to estimate threshold exceeding probabilities is proposed. To estimate the large-scale variability (spatial trend) of the process, the nonparametric local linear regression estimator, with the bandwidth selected by a method that takes the spatial dependence into account, is used. A bias-corrected nonparametric estimator of the variogram, obtained from the nonparametric residuals, is proposed to estimate the small-scale variability. Finally, a bootstrap algorithm is designed to estimate the unconditional probabilities of exceeding a threshold value at any location. The behavior of this approach is evaluated through simulation and with an application to a real data set.

\end{abstract}	


\section{Introduction}
\label{intro}

An important issue of many environmental studies is the estimation of the spatial uncertainty of certain variables of interest. Practical problems directly related with this subject are, for example, the control of pollution levels (e.g. pollutant concentration in soil, air, or water) or the prevention of damages caused by natural disasters. To tackle this problem, the usual procedure consists in applying a statistical method to try to estimate the probabilities of exceeding a given threshold, obtaining as a result a risk map. These maps provide important information for decision makers, and help them, for example, in the design of prevention policies to avoid adverse effects on human populations in areas where the estimated risk is high. 

Among the available geostatistical methods which may be applied to estimate exceeding probabilities, indicator kriging (IK) \citep[e.g.][]{Goovaerts1997}, disjunctive kriging (DK) \citep[see e.g.][]{Oliver1996} and Markov chain geostatistical modeling \citep[e.g.][]{Li2010} can be highlighted. IK is the most employed approach in this context. It is based on the application of the simple kriging linear predictor to indicator functions of the data \citep[see, e.g.][]{Journel83}. This method has some drawbacks, for instance, data discretization produces a loss of information, the estimated probabilities could be greater than one or negative,  and it could present order-relation problems \citep[see, e.g.][Section 6.3.3]{Chiles99}. To avoid these issues, \cite{tolosana2008} proposed the so-called simplicial indicator kriging. This method employs a simplex approach for compositional data to estimate  the conditional cumulative distribution function. This approach has also been adapted to the Bayesian framework in \cite{guardiola2011}. 
On the other hand, DK is also widely used as an alternative to IK \citep[see e.g.][]{Oliver1996}. This is a nonlinear estimation technique which usually assumes a Gaussian isofactorial model for the geostatistical process. However, as stated in \cite{Lark2004}, in general, there is no empirical evidence to recommend DK in preference to IK, or the opposite. 

The results obtained with these procedures could be unsatisfactory in practice due to the misspecification of the assumed parametric model. To deal with this concern, methods employing nonparametric estimation have also been proposed. For instance, \cite{garcia2012} suggest the use of a nonparametric kernel estimator of the indicator variogram in IK. 
In the spatio-temporal case, \cite{draghicescu2009} proposed the estimation of threshold exceedance probabilities by combining kernel smoothing in the time domain with spatial interpolation. \cite{cameletti2013} extended this approach, considering additional exogenous variables in the kriging interpolation and a block bootstrap algorithm to produce confidence regions for the risks.

It is important to stress that, given a spatial process $\left\{ Y(\mathbf{x}),\mathbf{x}\in D \subset \mathbb{R}^{d}\right\}$, the methods described above focus on the estimation of the conditional probability that the variable $Y$ exceeds a (critical) threshold value $c$ at a specific location $\mathbf{x}_{0}$, $P\left( Y(\mathbf{x}_{0})\geq c\left\vert \mathbf{Y}\right. \right)$, where $\mathbf{Y}=( Y(\mathbf{x}_{1}), \ldots ,Y(\mathbf{x}_{n}))^{t}$ denotes the vector of $n$ observed values at locations $\mathbf{x}_{1}, \ldots, \mathbf{x}_{n}$. They address this problem by predicting the indicator variable $I_{\left\{ Y(\mathbf{x}_{0})\geq c\right\} }$, 
since $E\left( I_{\left\{ Y(\mathbf{x}_{0})\geq c\right\} }\left\vert \mathbf{Y}\right. \right) = P\left( Y(\mathbf{x}_{0})\geq c\left\vert \mathbf{Y}\right. \right)$. 

This is the typical approach in geostatistics and could be the most appropriate in situations where the interest is a specific realization of the process, as for example in mining resource assessment or in weather forecast. However, in other cases, such as in climate studies, where the interest is to study the distribution of the process under certain conditions (specified through a trend), it could be preferable the estimation of the unconditional probability, also called long-term risk  \citep[see e.g][for additional commentaries about conditional and unconditional probabilities]{krz97, franks02}. This is the case of the present study.
We are interested in estimating the unconditional probability that the variable $Y$ exceeds a threshold $c$ at a location $\mathbf{x}_{0}$, that is:
\begin{equation}
r_{c}\left(\mathbf{x}_{0} \right)=P\left(Y(\mathbf{x}_{0})\geq c\right).
\label{rc}
\end{equation}
Note that the conditional probability can be very different from the unconditional one, because the distribution of $Y(\mathbf{x}_{0})\left\vert \mathbf{Y}\right.$ may have much less variability than the marginal distribution of $Y(\mathbf{x}_{0})$, especially in locations near the observed data. For that reason, if the target is the estimation of the unconditional probability, the geostatistical methods described above would not be appropriate. 

We assume the following general model:
\begin{equation}
Y(\mathbf{x})=m(\mathbf{x})+\varepsilon(\mathbf{x}),
\label{trendmodel}
\end{equation} 
where $m(\cdot)$ is the trend function, accounting for the large-scale variability, and the error term $\varepsilon$, representing the small-scale variability, is a second order stationary process with zero mean and covariogram $C(\mathbf{u}) = Cov(\varepsilon \left( \mathbf{x}\right) ,\varepsilon \left(\mathbf{x}+\mathbf{u}\right) )$. 

The geostatistical methods previously described are originally designed for stationary processes, although they could be adapted for the case of a non-constant trend. When the mean is not stationary, the traditional approach would be to consider a parametric model for the trend, obtain an estimate $\hat{m}(\cdot)$ from the data, and apply the desired method to the residuals $\hat{\varepsilon}_i=Y(\mathbf{x}_{i})-\hat{m}(\mathbf{x}_{i})$, $i=1, \ldots, n$. Nevertheless, it should be taken into account that the residual variability is different from that of the unobserved errors $\varepsilon(\mathbf{x}_{i})$, $i=1, \ldots, n$ \citep[see e.g.][Section 3.4.3]{cressie91}.

In this work, under the general spatial model (\ref{trendmodel}), and without assuming any parametric form for the trend function nor for the dependence structure of the process, a general nonparametric procedure to estimate the spatial risk (\ref{rc}) is proposed. This approach consists in estimating nonparametrically the trend function (with a bandwidth selected using an appropriate criterion). From the corresponding residuals, the variability of the process is nonparametrically modeled using a ``bias-corrected'' estimator similar to that described in \cite{Ruben2014}. Finally, a bootstrap algorithm designed to estimate the risk of exceeding a threshold value at a specific location is used. This procedure is a modification of the semiparametric bootstrap method described in \cite{mario11} in a seismological context.  The approach proposed in the present paper extends that method in two directions. First, trying to adequately reproduce the variability of the data, the bootstrap algorithm, apart from using the estimated variance-covariance matrix of the residuals (as in the former approach), also takes an estimate of the variance-covariance matrix of the errors into account. Second, instead of only using the estimated trend, simple kriging predictions of the residuals are also considered in the estimation of the exceeding probabilities. 

The organization of this paper is the following: Section \ref{sec2} reviews the nonparametric estimators for the mean function and for the variogram used in this work. At the end of this section, the proposed bootstrap algorithm is also described. In Section \ref{sec3}, some simulations and numerical results to illustrate the good performance of the method are included. An example of application to a real data set is shown in Section \ref{sec4}. Finally, Section \ref{conclu} summarizes the main conclusions.

\section{Methods}
\label{sec2}

In the proposed approach, the first step is the nonparametric estimation of the model components. The trend is estimated using the multivariate local linear estimator \citep[e.g.][]{Fan1996} and, from the corresponding residuals, the spatial dependence is estimated with an iterative algorithm to correct the biases due to the use of these residuals. Finally, these trend and covariogram estimates are employed in a bootstrap algorithm to approximate the exceeding probabilities.

\subsection{Local linear trend estimation and bandwidth selection for spatial data} 
\label{trendest}

From a nonparametric point of view, model (\ref{trendmodel}) has been studied by several authors, usually with the main interest of estimating the trend function $m(\cdot)$ under correlation. Some approaches used for this task include kernel-based methods, regression splines, wavelet techniques, and Fourier series expansions. In \cite{ops01}, a complete revision of these methods is presented. The derivation of asymptotic properties of the local linear estimator for the case of bidimensional correlated data and some methods for bandwidth selection in this situation are also provided in that paper.

In the spatial framework, the local linear estimator for $m(\cdot)$ at location $\mathbf{x}$ is obtained by solving for $\alpha$ the following least squares minimization problem:
$$\min\limits_{\alpha, \beta}\sum\limits_{i=1}^n \left\{Y(\mathbf{x}_i)-\alpha-\beta^{T}(\mathbf{x}_i-\mathbf{x}) \right\}^2  K_{\mathbf{H}}(\mathbf{x}_{i}-\mathbf{x}), $$
where  $K_{\mathbf{H}}(\mathbf{u})=\left\vert \mathbf{H} \right\vert ^{-1}K(\mathbf{H}^{-1}\mathbf{u})$, with $K$ being a multivariate kernel, and $\mathbf{H}$ a $d\times d$ nonsingular symmetric matrix. This estimator can be explicitly written as: 
\begin{equation}
 \hat{m}_{\mathbf{H}}(\mathbf{x})=\mathbf{e}_{1}^{t}\left( \mathbf{X}_{%
\mathbf{x}}^{t}{\mathbf{W}}_{\mathbf{x}}\mathbf{X}_{\mathbf{x}}\right) ^{-1}%
\mathbf{X}_{\mathbf{x}}^{t}{\mathbf{W}}_{\mathbf{x}}\mathbf{Y}\equiv {s}_{%
\mathbf{x}}^{t}\mathbf{Y},
\label{estimator}
\end{equation} 
where $\mathbf{e}_{1}$ is a vector with $1$ in the first entry and all other entries 0, $\mathbf{X}_{\mathbf{x}}$ is a matrix with $i$-th row equal to $(1,(\mathbf{x}_{i}-\mathbf{x})^{t})$, and
$$\mathbf{W}_{\mathbf{x}}=\mathtt{diag}\left\{ K_{\mathbf{H}}(\mathbf{x}_{1}-\mathbf{x}),\ldots ,K_{\mathbf{H}}(\mathbf{x}_{n}-\mathbf{x})\right\}.$$

The bandwidth matrix $\mathbf{H}$ controls the shape and size of the local neighborhood used to estimate the trend function $m(\mathbf{x})$. 
Geometrically, when the kernel function is spherically symmetric with bounded support, this neighborhood
is an ellipsis in $\mathbb{R}^2$ centered at $\mathbf{x}$. Only the observations within this neighborhood carry nonzero weights in the estimation. These weights depend on the spatial distances of the sample locations to $\mathbf{x}$.
Cross-validation (CV) techniques or plug-in methods, based on minimizing the asymptotic mean integrated squared error (AMISE), are frequently used as automatic bandwidth selection approaches in the case of independent data \citep[e.g.][]{Wand95}. However, it is well-known that when the data under study are dependent, the previous methods usually provide wrong smoothing parameters. For instance, in presence of positive correlated errors, the above procedures select bandwidths tending to undersmooth the trend estimates  \citep[see][]{liu01, ops01}.

In the case of correlated data and to avoid this problem, several modifications of the previous classical bandwidth selection methods have been proposed. We recommend the use of the ``bias corrected and estimated'' generalized cross-validation (CGCV) criterion, proposed in \cite{mario05}. This method selects the smoothing parameter $\mathbf{H}$ that minimizes:
\begin{equation}
CGCV(\mathbf{H}) = \frac{1}{n} \sum_{i=1}^{n}\left( \frac{Y(\mathbf{x}_{i}) - \hat{m}_{\mathbf{H}}(\mathbf{x}_{i})}{1-\frac{1}{n}tr\left(\mathbf{S\hat{R}}\right)}\right)^{2},
\label{gcvce}
\end{equation}
being $\mathbf{S}$ the $n\times n$ matrix whose $i$-th row is equal to $\mathbf{s}_{\mathbf{x}_{i}}^{t}$ (the smoother vector for $\mathbf{x} = \mathbf{x}_{i}$), $tr\left(\mathbf{S}\right)$ its corresponding trace, and $\mathbf{\hat{R}}$ is an estimator of the correlation matrix of the observations (incorporating the spatial dependence structure of the data). \cite{mario05} applied the method of the moments to obtain this estimate. However, in our approach, a nonparametric estimator of the covariogram function $C(\mathbf{u})$ will be used. Based on this estimate, the matrix $\mathbf{\hat{R}}$ can be obtained. A detailed description of this process is given in next section.

\subsection{Bias-corrected estimation of spatial dependence} 
\label{dep}

In geostatistics, the estimation of the spatial dependence is usually done through the semivariogram  $\gamma(\mathbf{u})=C(\mathbf{0})-C(\mathbf{u})$ \citep[see e.g.][]{cressie91}. In this work, local linear variogram estimation was considered due to its theoretical properties \citep{soidan} and its good performance in practice \citep{Ruben}.
When the trend is not stationary, the traditional approach consists in removing the trend and estimating the variogram from the residuals \citep[see][for the parametric case]{Neuman}. For instance, from the nonparametric residuals:
$$\hat{\mathbf{\varepsilon }}=\mathbf{Y}-\mathbf{SY},$$
where $\hat{\mathbf{\varepsilon }}=\left( \hat{\varepsilon}_{1},\ldots ,\hat{\varepsilon}_{n}\right) ^{t} $, a local linear estimate of the variogram $2\hat{\gamma}(\mathbf{u})$ is obtained as the solution for $\alpha$ of the least squares minimization problem: 
\begin{equation}
\min_{\alpha ,{\boldsymbol{\beta }}}\sum_{i<j}\left\{ \left( \hat{%
\varepsilon}_i - \hat{\varepsilon}_j \right)
^{2}-\alpha -{\boldsymbol{\beta }}^{t}\left( \mathbf{x}_{i}-\mathbf{x}_{j}-%
\mathbf{u}\right) \right\} ^{2}K_{\mathbf{G}}\left( \mathbf{x}_{i}-\mathbf{x}%
_{j}-\mathbf{u}\right) ,  
\label{svlinloc}
\end{equation}
where $\mathbf{G}$ is the corresponding bandwidth matrix. Following \cite{Ruben}, we propose to minimize the cross-validation relative squared error of the semivariogram estimates to select this bandwidth. 

No matter the method used to remove the trend, either parametric \citep[e.g.][Section 3.4.3]{cressie91} or nonparametric \citep[e.g.][]{Ruben2014}, the direct use of residuals in variogram estimation may produce a strong underestimation of the small-scale variability of the process. Simply note that:
\begin{equation*}
 Var(\hat{\mathbf{\varepsilon}})  = \boldsymbol{\Sigma}_{\hat{\mathbf{\varepsilon}}} = \boldsymbol{\Sigma} + \mathbf{B} ,
\end{equation*} 
where $\boldsymbol{\Sigma}$ is the covariance matrix of the errors, and $\mathbf{B} = \mathbf{S} \boldsymbol{\Sigma} \mathbf{S}^{t} - \boldsymbol{\Sigma} \mathbf{S}^{t} - \mathbf{S} \boldsymbol{\Sigma}$ is a square matrix representing the bias. 
A similar expression is obtained for the variogram:
\begin{equation}
Var(\hat{\mathbf{\varepsilon}}_i-\hat{\mathbf{\varepsilon}}_j)=Var({\mathbf{\varepsilon}}_i-{\mathbf{\varepsilon}}_j) + b_{ii}+b_{jj}-2b_{ij} ,
\label{correct_bias}
\end{equation} 
where $b_{ij}$ is the $(i,j)-$element of $\mathbf{B}$.

As the correct estimation of the small-scale variability is critical in the estimation of threshold-exceeding probabilities, a similar approach to that described in \cite{Ruben2014} was followed in this work. 
The first part of this approach consists in obtaining a bias-corrected pilot variogram estimate using an iterative algorithm. To do this, these steps are followed:
\begin{enumerate}
\item Using the residuals $\hat{\varepsilon}$ and applying (\ref{svlinloc}), obtain an uncorrected local linear estimate of the variogram.
\item Using the variogram estimate
obtained in the previous step, derive an approximation $\hat{\mathbf{B}}$ of $\mathbf{B}$.
\item Taking into account (\ref{correct_bias}), update the pilot variogram
estimate, replacing in (\ref{svlinloc}) the differences of 
the residuals $(\hat{\mathbf{\varepsilon}}_i-\hat{\mathbf{\varepsilon}}_j)^2$ by $(\hat{\mathbf{\varepsilon}}_i-\hat{\mathbf{\varepsilon}}_j)^2 - \hat{b}_{ii} - \hat{b}_{jj}+2 \hat{b}_{ij}$.
\item Repeat steps 2 and 3 until convergence.
\end{enumerate} 

In the original algorithm, \cite{Ruben2014} suggest the fit of a valid variogram model at step 2. However, in this case, $\mathbf{B}$ was directly approximated from pilot variogram estimates, using pseudo-covariances \citep[this algorithm is implemented in the function \texttt{np.svariso.corr} of the R package \texttt{npsp},][]{npsp}. Some empirical tests were performed and similar results were observed with both approaches, with a significant reduction in the computation time with this simpler approximation.
 The final variogram estimate was obtained by fitting a ``nonparametric'' isotropic Shapiro-Botha variogram model \citep{Shapiro1991} to the bias-corrected pilot estimate.

	To apply the previous algorithm, a nonparametric estimate of the trend is needed. However, an estimate of the variogram would be necessary to select an optimal bandwidth $\mathbf{H}$. To avoid this circular problem, the iterative procedure described in \cite{Ruben2014} was followed in this work. Starting with an initial bandwidth (e.g. obtained assuming independence), estimate the trend, compute the bias-corrected variogram estimator, fit a valid variogram model, select the bandwidth using the CGCV criterion (\ref{gcvce}), and so forth. In practice, one iteration of this procedure is usually sufficient.

\subsection{Bootstrap algorithm} 
\label{boot}

In this section, a bootstrap algorithm designed to adequately reproduce the variability of the data is proposed. This  method is a modification of the semiparametric bootstrap described in \cite{mario11}. The specific steps of the new approach are the following:

\begin{enumerate}

\item Obtain estimates of both $\boldsymbol{\Sigma}_{\hat{\mathbf{\varepsilon}}}$ and $\boldsymbol{\Sigma}$, and the corresponding Cholesky factorizations:

\begin{enumerate}

\item Select a bandwidth matrix $\mathbf{H}$ to estimate the trend (for instance, using the algorithm described in previous section).

\item Compute the trend estimates $\hat{m}_{\mathbf{H}}(\cdot)$, given by (\ref{estimator}), and the respective residuals $\hat{\varepsilon}_{i} = Y(\mathbf{x}_i) - \hat{m}_{\mathbf{H}}(\mathbf{x}_i)$, for $i=1,\ldots, n$.

\item Obtain an estimate of the variogram of the residuals, computing the local linear pilot estimates (\ref{svlinloc}) and fitting a valid Shapiro-Botha model to these estimates. Obtain the corresponding (estimated) covariance matrix $\hat{\boldsymbol{\Sigma}}_{\hat{\mathbf{\varepsilon}}}$ and its Cholesky decomposition $\hat{\boldsymbol{\Sigma}}_{\hat{\mathbf{\varepsilon}}}=\mathbf{L}_{\hat{\mathbf{\varepsilon}}}\mathbf{L}_{\hat{\mathbf{\varepsilon}}}^{t}$.

\item Using the procedure described in previous section, obtain a bias-corrected estimate of the variogram. Compute the corresponding (estimated) covariance matrix $\hat{\boldsymbol{\Sigma}}$ and its Cholesky decomposition $\hat{\boldsymbol{\Sigma}}=\mathbf{LL}^{t}$.
\end{enumerate}

\item Generate bootstrap samples with the estimated spatial trend $\hat{m}_{\mathbf{H}}(\mathbf{x}_{i})$ and adding bootstrap errors generated as a spatially correlated set of errors. The bootstrap errors are obtained as follows:

\begin{enumerate}

\item Compute the ``independent'' residuals $\mathbf{e}=\mathbf{L}^{-1}_{\hat{\mathbf{\varepsilon}}}\hat{\mathbf{\varepsilon }}$, where $\mathbf{e}=\left( e_1,\ldots ,e_{n}\right) ^{t}$.

\item These variables are centered and, from them, obtain an
independent bootstrap sample of size $n$, denoted by $\mathbf{e}^{\ast}=\left( e_{1}^{\ast },\ldots ,e_{n}^{\ast
}\right) ^{t} $.

\item Next, compute the bootstrap errors $\mathbf{\varepsilon }^{\ast} = \mathbf{Le}^{\ast }$, where $\mathbf{\varepsilon }^{\ast}=\left( {\varepsilon}_{1}^{\ast },\ldots ,{\varepsilon}_{n}^{\ast}\right) ^{t}$, and obtain the bootstrap resamples of the spatial process $Y^{\ast }(\mathbf{x}_{i})=\hat{m}_{\mathbf{H%
}}(\mathbf{x}_{i})+{\varepsilon}_{i}^{\ast },\text{ }i=1,2,\ldots ,n.$ 

\end{enumerate}

\item Compute the kriging prediction $\hat{Y}^{\ast}(\mathbf{x}_{0})$  at each unsampled location $\mathbf{x}_{0}$ from the bootstrap sample $\left\{ Y^{\ast}(\mathbf{x}_1),\ldots,Y^{\ast}(\mathbf{x}_n)\right\}$ (applying the nonparametric local linear regression estimator to the bootstrap sample, using the same bandwidth $\mathbf{H}$ as for the original analysis, and adding the simple kriging predictions obtained from the corresponding residuals).  

\item Repeat steps 2 and 3 a large number of times $B$ (in our analysis,  $B=1,000$).
Therefore, for each unsampled location $\mathbf{x}_{0}$, $B$ bootstrap replications $\hat{Y}^{\ast(1)}(\mathbf{x}_{0}), \ldots, \hat{Y}^{\ast(B)}(\mathbf{x}_{0})$ are obtained.

\item Finally, a map with estimates of the unconditional exceeding probabilities, given in (\ref{rc}), is obtained by calculating the relative frequencies across bootstrap replicates, 
\begin{equation}
\hat{r}_{c}\left(\mathbf{x}_0 \right)=\frac{1}{B}\sum_{j=1}^{B} I_{\left\{ \hat{Y}^{\ast(j)}(\mathbf{x}_{0})\geq c\right\} },
\label{r_est}
\end{equation}
of how often each location $\mathbf{x}_{0}$ is included in the at-risk area.

\end{enumerate}

Note that, on the contrary of what happens in conditional probability estimation, the bootstrap samples generated at the end of the step 2 will not coincide, in general, with the observed values.
Moreover, this method could be adapted to the construction of confidence (prediction) intervals or hypothesis testing. Indeed, with the corresponding adjustments, it can be applied in the case of independent data to correct the bias that arise when classical residual bootstrap algorithms are used.

Alternatively, especially if the goal is only to obtain risk maps, response values could be generated at prediction locations rather than at data locations (and eliminating step 3 in the previous algorithm). In this way, the process variability would be expected to be better reproduced. Nevertheless, very similar results were obtained in practice with both approaches. One advantage of the proposed algorithm is that it allows to simultaneously perform inferences about characteristics of interest of the process (for instance, about the trend and the variogram).

\section{Simulation results}
\label{sec3}

In this section, the practical behavior of the proposed approach is analyzed by simulations under different scenarios. Sample sizes of $n=10\times 10$, $17\times 17$ and $20\times 20$ were considered in the study. For each scenario, $N=1,000$ samples were generated on a regular grid in the unit square $D = [0,1] \times [0,1]$ following model (\ref{trendmodel}), with mean function  
$$m(x_{1},x_{2})=2.5 + \sin(2\pi x_{1})+4(x_{2}-0.5)^{2},$$ 
and random errors $\varepsilon _{i}$ normally distributed with zero mean and isotropic exponential covariogram 
\begin{equation}
\gamma _{\theta }(\mathbf{u})=c_{0}+c_{1}\left( 1-\exp \left( -3\frac{\Vert 
\mathbf{u}\Vert }{r}\right) \right) ,
\label{covar}
\end{equation} 
(for $\mathbf{u}\neq \mathbf{0}$), where $c_{0}$ is the nugget effect, $c_{1}$ is the partial sill ($\sigma^{2} = c_0 + c_1$ is the sill or variance), and $r$ is the practical range. Several degrees of spatial dependence were studied, considering values of $r=0.25$, $0.50$, and $0.75$, $\sigma^{2}=0.16$, $0.32$, and $0.64$, and nugget values of 0\%, 25\% and 50\% of $\sigma^{2}$. 

For each parameter configuration, given a threshold $c$, the theoretical probabilities $r_{c}\left(\mathbf{x}_0 \right)$, given in (\ref{rc}), were computed in a regular $50 \times 50$ grid. In each simulation, estimates of these probabilities $\hat{r}_{c}\left(\mathbf{x}_0 \right)$, given in (\ref{r_est}), were also obtained with the proposed approach, using $B=1,000$ bootstrap replications. Threshold values of $c=2.0, 2.5, 3.0$ and $3.5$ were considered in this analysis.

The trend was estimated using the local linear estimator (\ref{estimator}) with the multiplicative triweight kernel. To reduce computation time and to avoid the effect of the trend bandwidth selection criterion on the results, the bandwidth minimizing the mean average squared error:
$${\rm MASE}(\mathbf{H})=\frac{1}{n}(\mathbf{Sm}-\mathbf{m})^t (\mathbf{Sm}-\mathbf{m})+\frac{1}{n}tr(\mathbf{S}\mathbf{\Sigma}\mathbf{S}^t),$$
where $\mathbf{m}=(m(\mathbf{x_1}), \ldots , m(\mathbf{x_n}))^t$, was employed. This optimal bandwidth cannot be used in practice, in which case we recommend the use of the CGCV criterion (\ref{gcvce}) (similar results were observed  with both criteria in preliminary simulation experiments). 

Regarding the variogram, the (uncorrected) variogram estimates (\ref{svlinloc}) and the bias-corrected version were computed on a regular grid up to the 55\% of the largest sample distance. The bandwidth matrices were selected applying the cross-validation relative squared error criterion. Finally, isotropic Shapiro-Botha variogram models were fitted to these estimates.

To study the effect of the covariance matrix estimation method in the performance of the bootstrap algorithm, the proposed complete procedure (which uses a bias-corrected variogram estimate and is denoted by ``corrected'' in the results) was compared with the approaches using, respectively, the true covariance matrix $\boldsymbol{\Sigma}$ (denoted by ``theoretical''), and the estimated covariance matrix of the residuals $\hat{\boldsymbol{\Sigma}}_{\hat{\mathbf{\varepsilon}}}$ (computed from the uncorrected pilot local linear variogram estimates and denoted by ``residual''). To validate the performance of the different approaches, the squared errors:
$$
{\rm SE}\left(\mathbf{x} \right)=(r_{c}\left(\mathbf{x} \right)  - \hat{r}_{c}\left(\mathbf{x} \right) )^2 ,
$$
were computed in the estimation grid.

For reasons of space, only some representative results are shown here. For instance, the mean, median and standard deviations of the squared errors ($\times 10^{-2}$), for $c=2.5$, $\sigma ^{2}=0.16$, $r=0.5$ and $c_{0}=0.04$, are presented in Table \ref{t1}. As expected, the best results were obtained when the true covariance matrix $\boldsymbol{\Sigma}$ was used in the procedure, with the proposed approach providing very close results, specially considering the median of the squared errors. It is also observed that the bias-correction in variogram estimation significantly improved the results. Average squared error reductions of about 40\% of the ``corrected'' proposal with respect to the ``residual'' version were observed in the different scenarios. Additionally, to check the influence of using simple kriging predictions of the residuals, the results obtained with the ``residual'' version could be compared with those obtained in \cite{mario11}, observing a significant improvement with the residual version of the proposed approach (leading to a reduction of about 25\% in the averaged squared errors).

\begin{table}[htb]
\caption{Mean, median and standard deviation of the squared errors ($\times 10^{-2}$) of the theoretical, residual and corrected versions of the proposed procedure, for threshold $c=2.5$, $\sigma ^{2}=0.16$, $r=0.5$ and $c_{0}=0.04$. Sample sizes $n=10 \times 10$, $17 \times 17$ and $20\times 20$ are considered.}
\begin{center}
\begin{tabular}{l|ccc|ccc|ccc}
\hline
 & \multicolumn{3}{|c}{$n=10 \times 10$} & \multicolumn{3}{|c}{$n=17 \times 17$} & \multicolumn{3}{|c}{$n=20 \times 20$} \\ \hline
 & mean & median & sd & mean & median & sd & mean & median & sd  \\ \hline
Theoretical	& 2.30 & 0.09 &  5.40 & 2.05 & 0.08 & 4.97 & 1.90 & 0.08 & 4.65 \\ 
Residual 	& 5.00 & 0.16 & 11.00 & 4.10 & 0.16 & 9.40 & 3.80 & 0.16 & 8.60 \\ 
Corrected 	& 2.60 & 0.09 &  6.40 & 2.40 & 0.09 & 5.80 & 2.20 & 0.08 & 5.40 \\ \hline
\end{tabular}
\end{center}
\label{t1}
\end{table}

\begin{figure}[!htb]
\begin{center}
\begin{tabular}{cc}
(a) & (b) \\
{\includegraphics[height=4cm]{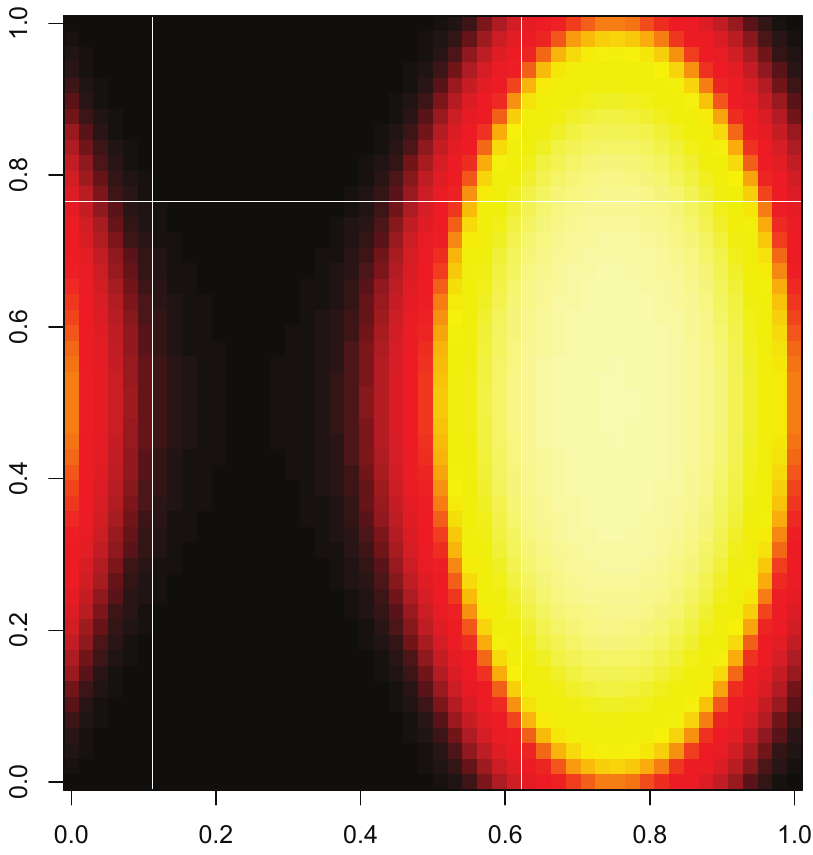}} & {\includegraphics[height=4cm]{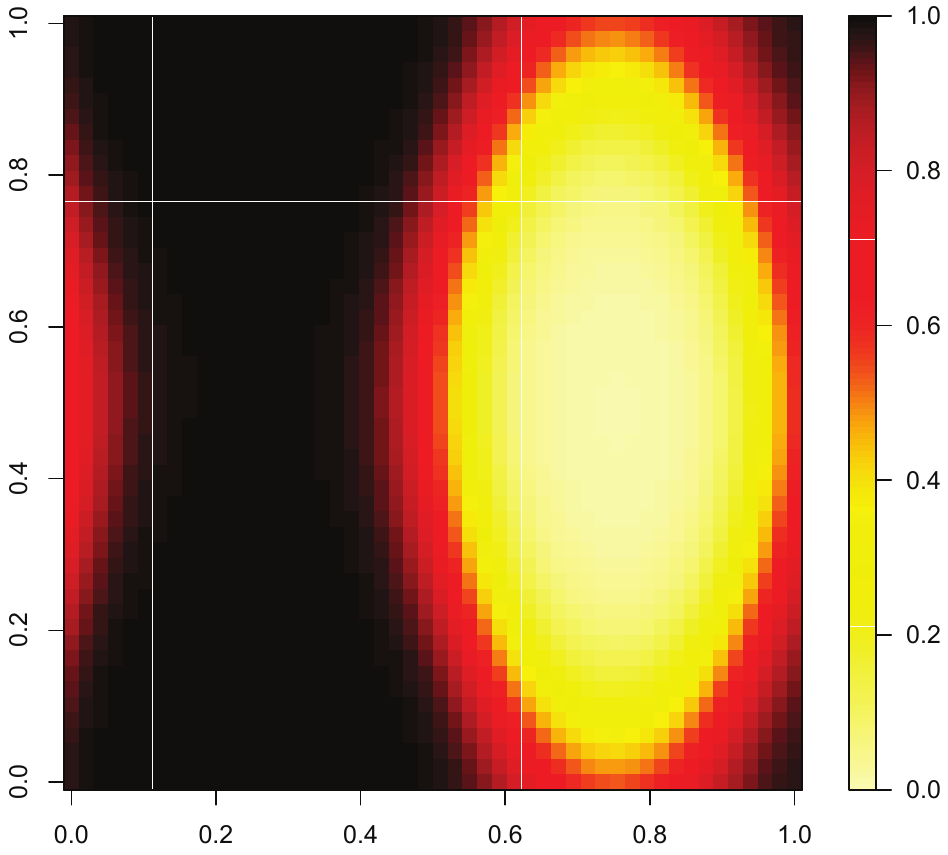}} \\
\end{tabular}
\caption{Probabilities of exceeding threshold 2.5 for $n = 20\times 20$, $\sigma ^{2}=0.16$, $r=0.5$ and $c_{0}=0.04$: (a) Theoretical probabilities, and (b) averaged (across simulations) estimated probabilities using the ``corrected'' estimator.}
   \label{fig1}   
\end{center}
\end{figure}

The good performance of the proposed procedure can also be visually observed  in Figure \ref{fig1}, where the theoretical probabilities $r_{c}\left(\mathbf{x}_0 \right)$, for $c=2.5$, are shown in left panel and, in right panel, the corresponding averages (over the $N$ replicas) of the estimated probabilities $\hat{r}_{c}\left(\mathbf{x}_0 \right)$ obtained with the ``corrected'' approach.   

\begin{figure}[!htb]
\begin{center}
\begin{tabular}{ccc}
(a) & (b) & (c)\\
{\includegraphics[height=3.5cm]{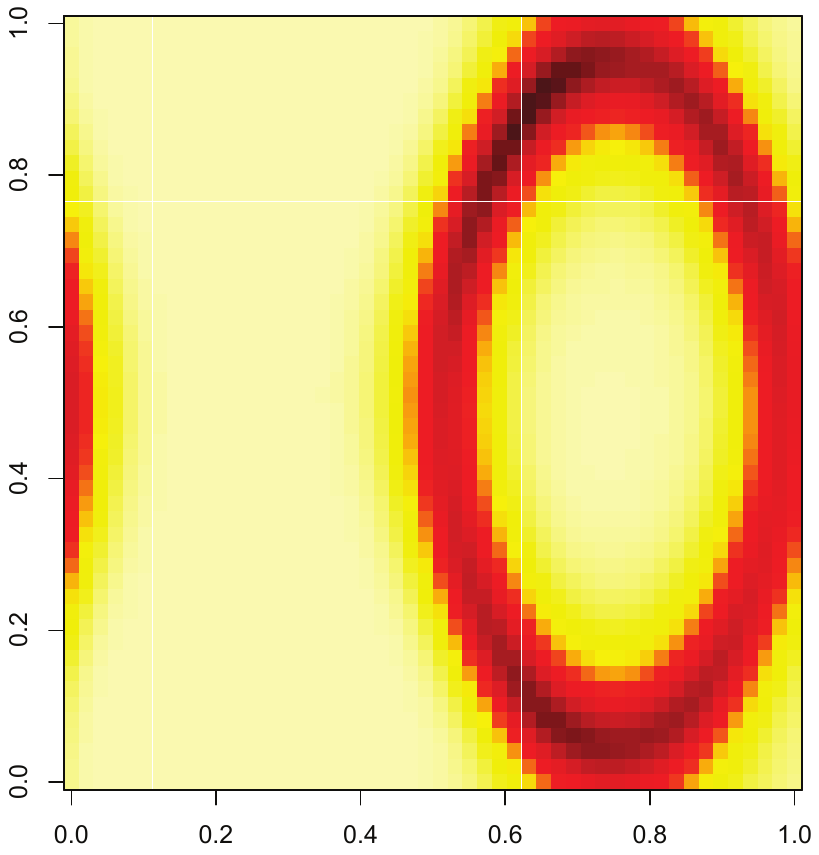}} & {\includegraphics[height=3.5cm]{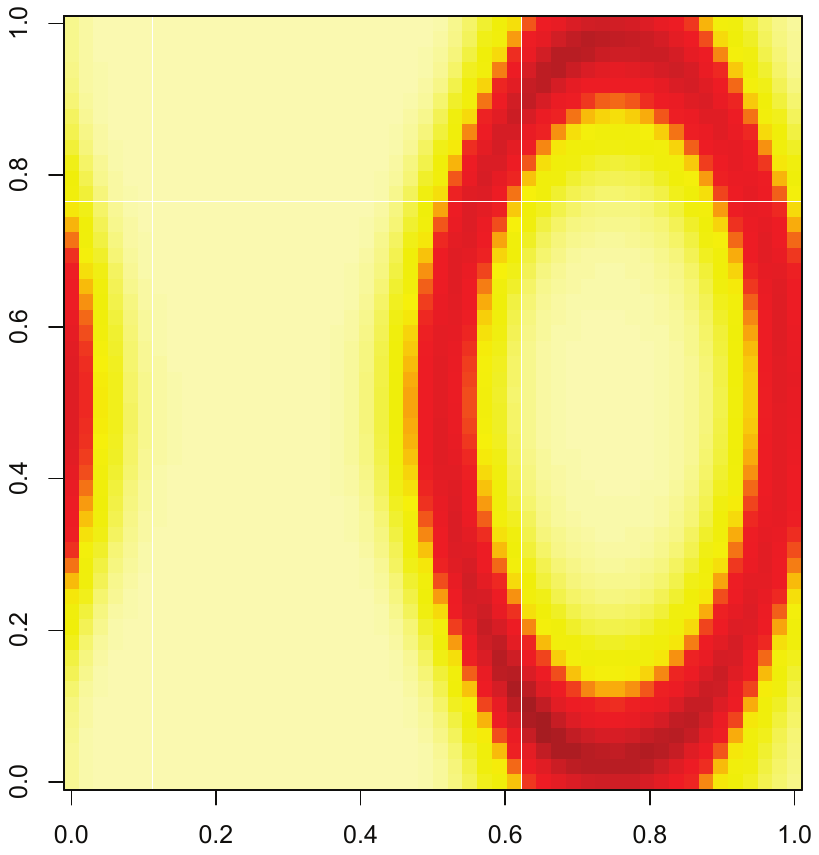}} & {\includegraphics[height=3.5cm]{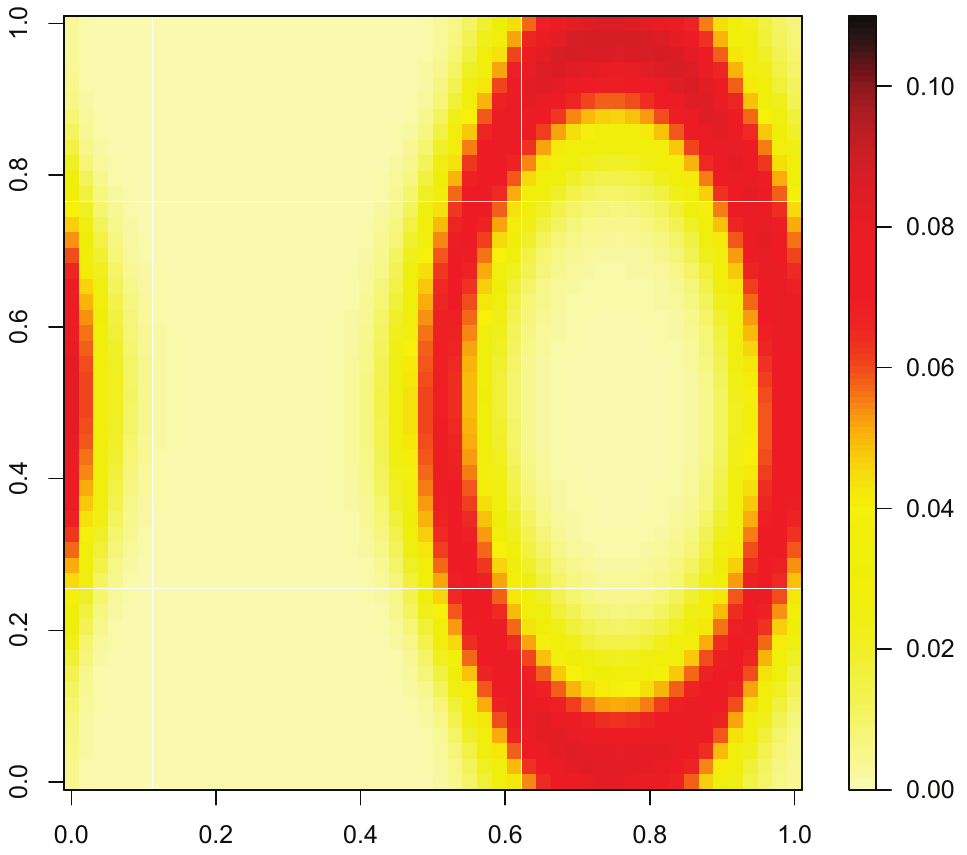}} \\
\end{tabular}
   \caption{Averaged squared errors surfaces using the ``corrected'' estimator for $c=2.5$, $\sigma ^{2}=0.16$, $r=0.5$, $c_{0}=0.04$, and different sample sizes ($n=10 \times 10$ in left panel, $n=17 \times 17$ in central panel, and $n=20 \times 20$ in right panel).}
   \label{fig2}
\end{center}
\end{figure}

Figure \ref{fig2} shows the averaged squared errors of the estimates obtained with the ``corrected'' procedure, for $c=2.5$, $\sigma ^{2}=0.16$, $r=0.5$, $c_{0}=0.04$, and the different sample sizes. Consistency of this algorithm can be presumed from these results (see also Table \ref{t1}). Observing the corresponding surfaces of the trend estimation bias (not shown for reason of space), it seems that this bias plays an important role in the asymptotic behavior of the proposed approach (in the scenarios considered here).

Averaged squared errors of the proposed method for different spatial dependence degrees are shown in Table \ref{t2} (for $c=2.5$, $\sigma ^{2}=0.16$, $n=20 \times 20$, and the different values of nugget and practical range). As expected a better estimation precision was observed when the spatial dependence is weak 
\cite[either when the nugget value increases or the range parameter decreases, smaller averaged squared errors were obtained; see e.g.][Section 1.3., for general comments on estimation under dependence]{cressie91}.

\begin{table}[htb]
\caption{Averaged of squared errors ($\times 10^{-2}$) of the estimated probabilities with the corrected estimator for $c=2.5$, $\sigma ^{2}=0.16$, $n=20 \times 20$, and the different nuggets and practical ranges.}
\begin{center}
\begin{tabular}{l|cc|cc|cc}
\hline
 & \multicolumn{2}{|c}{$c_0=0\%$} & \multicolumn{2}{|c}{$c_0=25\%$}& \multicolumn{2}{|c}{$c_0=50\%$} \\ \hline
 & Residual & Corrected & Residual & Corrected & Residual & Corrected \\ \hline
$r=0.25$ & 2.90 & 1.60 & 3.00 & 1.60 & 3.00 & 1.80  \\ 
$r=0.50$ & 3.80 & 2.34 & 3.80 & 2.20 & 3.70 & 2.25  \\ 
$r=0.75$ & 4.50 & 2.90 & 4.40 & 2.70 & 4.10 & 2.70 \\ \hline
\end{tabular}
\end{center}
\label{t2}
\end{table}

Finally, we have completed the simulation study performing some experiments with non-regular sample designs. The results obtained follow the same line as those with regular sampling and similar conclusions could be deduced from them.
As an example, Table \ref{t3} shows similar results to those presented in Table \ref{t1}, but considering samples with locations generated by a bidimensional uniform distribution over the unit square. Similar results were obtained with other sampling designs.
It is important to remark that, in the case of non-regular designs, the computational cost is larger. In the fixed design case, the optimal bandwidth matrices (for trend and variogram estimation) and the smoothing matrix $\mathbf{S}$ only need to be computed once, whereas a random design will require the computation of these matrices at each iteration, considerably increasing the computational time.

\begin{table}[htb]
\caption{Mean, median and standard deviation of the squared errors ($\times 10^{-2}$) of the theoretical, residual and corrected versions of the proposed procedure with non-regular sampling (uniform distribution in the unit square), for threshold $c=2.5$, $\sigma ^{2}=0.16$, $r=0.5$ and $c_{0}=0.04$. Sample sizes $n=10 \times 10$, $17 \times 17$ and $20\times 20$ are considered.}
\begin{center}
\begin{tabular}{l|ccc|ccc|ccc}
\hline
 & \multicolumn{3}{|c}{$n=10 \times 10$} & \multicolumn{3}{|c}{$n=17 \times 17$} & \multicolumn{3}{|c}{$n=20 \times 20$} \\ \hline
 & mean & median & sd & mean & median & sd & mean & median & sd  \\ \hline
Theoretical 	& 2.50	& 0.13 &	6.00 &	2.20 &	0.09 &	5.20 &	2.10 &	0.08 &	5.10 \\
Residual 	& 4.50 &	0.17 &	10.00 &	4.52 &	0.21 &	10.70 &	6.20 &	0.34 &	14.00 \\
Corrected 	& 2.70 &	0.17 &	6.50 &	2.30 &	0.10 &	5.60 &	2.28 &	0.09 &	5.63 \\ \hline
\end{tabular}
\end{center}
\label{t3}
\end{table}

\section{Application to real data}
\label{sec4}

\begin{figure}[!htbp]
\begin{center}
\begin{tabular}{cc}
(a) & (b) \\
{\includegraphics[height=3cm]{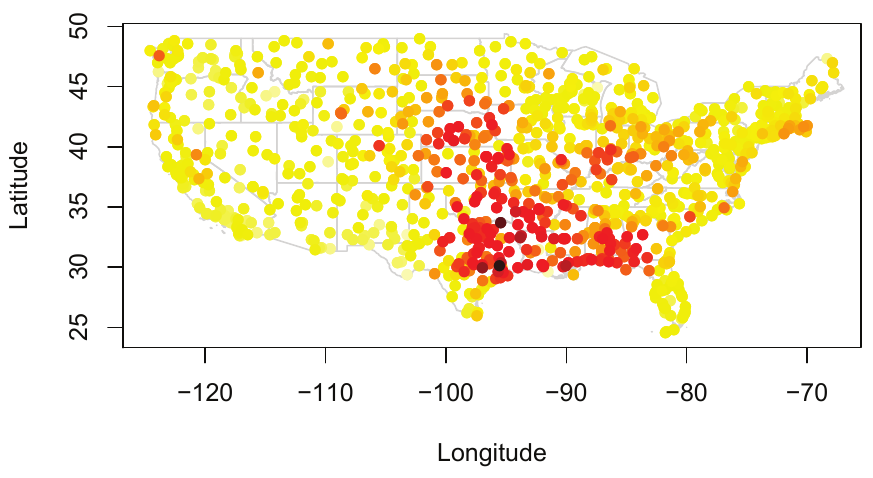}} & {\includegraphics[height=3cm]{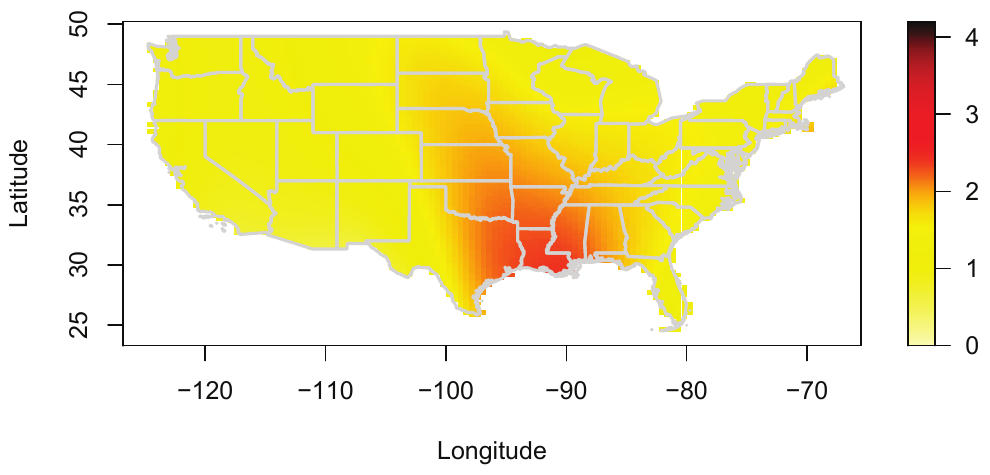}} \\
(c) & (d) \\
{\includegraphics[height=3cm]{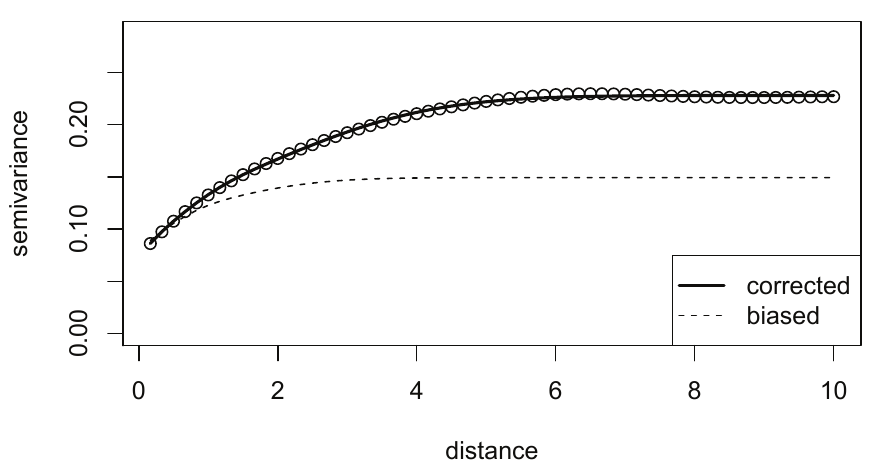}} & {\hspace{-0.76cm}\includegraphics[height=3cm]{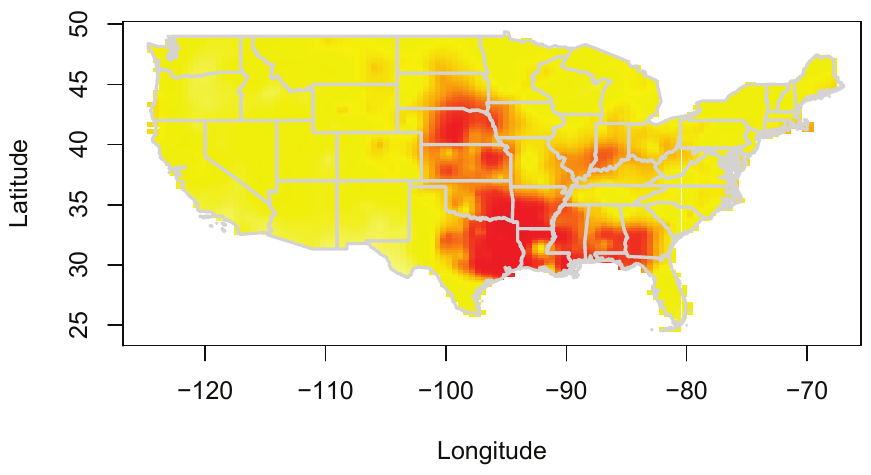}} \\
\end{tabular}
\caption{(a) Spatial locations and observed values, (b) final nonparametric trend estimates, (c) nonparametric corrected and uncorrected semivariogram estimates, and (d) kriging predictions; corresponding to total precipitations in March 2016 (square-root of rainfall inches).}
\label{fig3}
\end{center}
\end{figure}

In this section, as an example, the proposed methodology was applied to climatological data. The data set consists of total precipitations (rainfall inches) during March 2016 recorded over 1053 locations on the continental part of USA (available at \textit{http://www.ncdc.noaa.gov/pclcd}). After an initial descriptive analysis, the response variable was root-transformed to achieve symmetry. The spatial distribution of these data is shown in Figure \ref{fig3}a.

As described at the end of Section \ref{dep}, an iterative process to estimate the trend and the variogram was used. In this case, only two iterations were needed. First, the traditional CV criterion was employed to obtain a pilot bandwidth $\mathbf{H}_0$. This bandwidth was used to estimate the trend and to obtain the corresponding residuals. Analyzing these residuals, the assumption of isotropy was found plausible. Therefore, pilot local linear isotropic semivariogram estimates were computed, minimizing the cross-validation relative squared error to select the bandwidth. To correct the existing bias, a bias-corrected pilot variogram estimate was obtained using the algorithm described in Section \ref{dep} \citep[using the \texttt{np.svariso.corr} function of the R package \texttt{npsp},][]{npsp}. The variogram estimate at this step was obtained by fitting an isotropic Shapiro-Botha model to these pilot estimates. In the next step, using the unbiased variogram estimate previously obtained, a new bandwidth $\mathbf{H}_1$ was selected applying the CGCV criterion (\ref{gcvce}). A new nonparametric trend estimate was computed and the variogram was re-estimated from the current residuals. This procedure should be repeated until convergence. However, as the variogram estimates obtained in the first and second steps were very similar, no additional iterations were needed in this case. The final trend estimates are shown in Figure \ref{fig3}b. Figure \ref{fig3}c displays the (uncorrected) initial and the final variogram estimates. As expected, the direct use of uncorrected residuals produces an underestimation of the spatial variability of the process and may have a significant impact on the estimated risk maps. Finally, in Figure \ref{fig3}d, the kriging predictions obtained with the final trend and variogram estimates are shown.

\begin{figure}[!htb]
\begin{center}
\begin{tabular}{cc}
(a) & (b) \\
{\includegraphics[height=3cm]{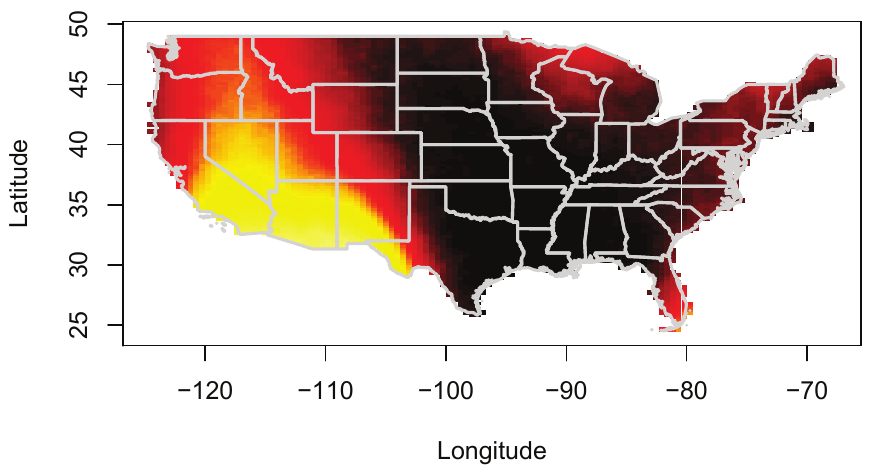}} & {\includegraphics[height=3cm]{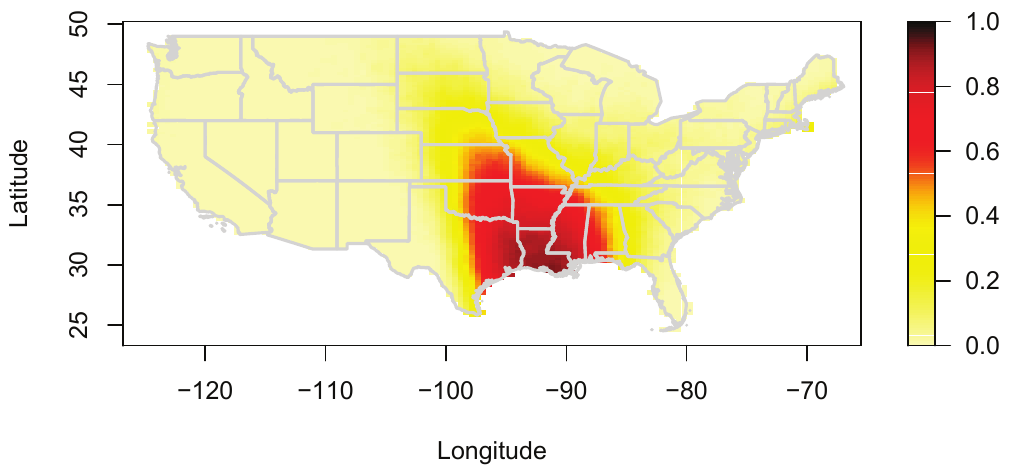}} \\
\end{tabular}
\caption{Maps with the (unconditional) probabilities, $\hat{r}_{c}\left(\mathbf{x} \right)$, of occurring a total precipitation larger than or equal to the threshold (a) $c=1.0$ and (b) $c=2.0$ (square-root of rainfall inches) in the area of study. }   
	 \label{fig4}
\end{center}
\end{figure}

Using the bootstrap procedure described in Section \ref{boot}, pointwise probabilities of exceeding different threshold values $c$ were estimated in the region of study. For instance, Figure \ref{fig4} shows maps with the (unconditional) estimated probabilities $\hat{r}_{c}\left(\mathbf{x} \right)$, for $c =1.0$ and $2.0$ (1.0 and 4.0 rainfall-inches, respectively). 
These maps show an area with low probability of precipitation, mostly corresponding to the States of Arizona, Utah and New Mexico. These regions of the western United States are characterized for being dry areas with semiarid climate.
Moreover, the results also show a zone with high probability of heavy rains corresponding to several southern US states. It is important to note that for the period considered, heavy rains produced floods in those areas, especially along
the Sabine River, located between the States of Texas and Louisiana.

This is a simple illustration of the results that can be obtained with this approach. This kind of probability maps could be useful, for instance, in long-term risk management to prevent damage caused by floods or in the assessment of crop production. However, in certain situations, more sophisticated models could be required to address these type of problems properly, for example, considering additional explanatory variables or a temporal component.

\section{Conclusions}
\label{conclu}

In this paper, a bootstrap algorithm to estimate risk maps for geostatistical data is proposed. The probability of obtaining a value of a response variable larger than or equal to a particular threshold, at a specific location, is estimated with this approach. A general model, considering that the variable of interest can be written as the sum of two terms, a deterministic trend function accounting for the large-scale variability and an error random process representing the small-scale variability, is assumed. As part of the general method proposed, nonparametric estimators of the trend function (the local linear estimator) and the variability \citep[a slight modification of the ``bias-corrected'' method proposed in][]{Ruben2014} are considered. These nonparametric approaches produce more flexible estimations of the trend and the dependence of the data, and problems due to model misspecification are avoided. This approach can be easily adapted to the construction of confidence or prediction intervals and to hypothesis testing, and even to the case of independent data, which shows the versatility of the method.

The performance of the new proposal was validated through a comprehensive simulation study, showing its good behavior under different scenarios, including several degrees of spatial dependence. The simulation results confirm that correcting the bias due to the direct use of residuals in variogram estimation, as well as adding the kriging predictions of the residuals in the bootstrap method, may have a significant improvement on the risk assessment procedure. 

As stated in the Introduction, the results obtained with the proposed method should not be compared with those obtained using traditional geostatistical approaches (such as indicator or disjunctive kriging). The aim of those conventional procedures is the estimation of the conditional probability, while the method proposed here tries to approximate the unconditional risk. 
However, it could be properly modified to estimate the conditional probability. The general idea would be to perform a conditional bootstrap, combining a resampling algorithm similar to that described in Section \ref{boot}, with kriging predictions \citep[see e.g.][Section 3.2.6, for details on conditional simulation]{cressie91}. 

The procedures used in this study were implemented in the statistical environment R, using the functions for nonparametric trend and variogram estimation supplied with the \texttt{npsp} package \citep[][available on CRAN]{npsp}.

\section*{Acknowledgements}\label{acknowledgements}
The research of Rub\'{e}n Fern\'{a}ndez-Casal and Mario Francisco-Fer\-n\'{a}n\-dez has been partially supported by the Conseller\'ia de Cultura, Educaci\'on e Ordenaci\'on Universitaria of the Xunta de Galicia through the agreement for the Singular Research
Center CITIC, and by Grant MTM2014-52876-R. The research of Sergio Castillo has been partially supported by the Universidad de las Fuerzas Armadas ESPE, from Ecuador. 

%

\end{document}